\documentclass[aps,prl,twocolumn,showpacs,unsortedaddress]{revtex4}
\usepackage{color,graphicx}
\usepackage{amsmath,amssymb}
\usepackage[colorlinks,citecolor=red]{hyperref}
\begin{document}
\let\up=\uparrow
\let\down=\downarrow
\let\dag=\dagger
\newcommand\unit[1]{\operatorname{#1}}
\newcommand\half{\frac{1}{2}}
\newcommand\calL{\mathcal{L}}

\title{Coupling Circuit Resonators Among Themselves and To Nitrogen-Vacancy
  Centers in Diamond}

\author{Myung-Joong Hwang}
\affiliation{Department of Physics, Pohang University of Science and
  Technology, Pohang 790-784, Korea}

\author{Mahn-Soo Choi}
\email{choims@korea.ac.kr}

\affiliation{Department of Physics, Korea University, Seoul 136-713, Korea}
\affiliation{Asia Pacific Center for Theoretical Physics, Pohang 790-784,
  Korea}

\begin{abstract}
We propose a scheme to couple NV centers in diamond through
coplanar waveguide resonators.  The central conductor of the resonator is
split into several pieces which
are coupled strongly with each other via simple capacitive junctions or
superconducting Josephson junctions.  The NV centers are then put at the
junctions.  The discontinuity at the junctions induces a large local magnetic
field, with which the NV centers are strongly coupled to the circuit
resonator.  The coupling strength $g$ between the resonator and the NV center
is of order of $g/2\pi\sim 1$--$30\unit{MHz}$.
\end{abstract}

 \maketitle

A nitrogen-vacancy (NV) center in diamond consists of a nitrogen
atom substituting a carbon atom and a vacancy trapped
adjacent to the substitutional nitrogen.  In its ground state, the
negatively charged NV center has a spin triplet, which is separated
by optical transitions from the excited states. The long spin
coherence time and fairly easy optical initialization and read out
of the ground spin state make the NV center an excellent candidate
for quantum information processor and quantum information
storage~\cite{Wrachtrup06a}. The coherent manipulation of the single
spin or multiple spins of electrons and nuclei within a single NV
center~\cite{Neumann08a,Nizovtsev05a,Childress06a} or locally
interacting NV centers~\cite{Neumann10a} have been demonstrated
experimentally. To build a scalable quantum
information processors (QIP), however, controlled coupling between
distant NV centers is yet to be achieved.

A superconducting coplanar waveguide resonator, defined by a single
centimeters-long central conductor 
between two ground
half-planes, has been successfully used as a quantum bus for
charge-based superconducting qubits~\cite{Majer07a,DiCarlo09a} by
exploiting a strong \emph{electric} dipole
coupling~\cite{Blais04a,Wallraff04a}. Building a hybrid quantum
device using the superconducting resonator as a quantum bus for the
spin qubits (including NV centers) is desirable for the scalable
QIP, because then we can take advantage of the scalability and low
dissipation of the circuit-QED system as well as the long coherence
time of spin qubits. However, it has been limited only to collective
excitation of the spin ensemble due to a small \emph{magnetic}
dipole coupling of a single spin qubit with the resonator of an
order of $10$
Hz~\cite{Imamoglu09a,Verdu09a,Wesenberg09a,Kubo10a,Schuster10a}.
Using a flux qubit as a mediator between the NV center and the
resonator is suggested to enhance the coupling strength recently,
but then the short coherence time of the flux qubit limits the
coherence time of the entire system~\cite{Twamley10a}.

In this paper, we describe how to realize a quantum bus for distant
NV centers using a series of superconducting resonators coupled by
either Josephson or capacitive junctions. We note that inserting a
Josephson junction into the central conductor of the resonator
enhances \emph{local} magnetic field by a factor of
$10^4$~\cite{Bourassa09a,Niemczyk10a}. Interestingly, a simple
capacitive junction can also enhance the local magnetic field by the
same factor as demonstrated below, which would allow us to
circumvent the difficulties of fabricating many Josephson junctions
in the central conductor. By putting single NV center at each
junction, the magnetic dipole coupling strength between the
resonator and the NV centers reaches $g/2\pi\sim1$--$30\unit{MHz}$.
The interaction between NV centers can be achieved by exchanging
virtual photons and can be turned on and off by bringing them in and
out of resonance with the resonator frequency by the external
magnetic field. Thus quantum gate operations can be performed in the
same manner as the circuit-QED system~\cite{Majer07a,DiCarlo09a}.
An important difference is that the NV center qubit can be measured
optically instead of being measured dispersively through the cavity.
It means the high-loss cavity used for the fast measurement in the
circuit-QED experiments~\cite{Majer07a,DiCarlo09a} is no longer
necessary. We also note that our scheme offers orders of magnitude
stronger coupling strength than recently proposed nanomechanical
resonators-based quantum bus~\cite{Rabl10a,Chen10a} for the NV
centers as well as schemes that use the flux
qubits~\cite{Twamley10a,Marcos10a}, which enables us to perform fast
quantum gate operations. Moreover, recent experimental demonstration
of the coupling between an ensemble of NV centers and the
superconducting resonator~\cite{Kubo10a} along with experiments
realizing the superconducting resonator with the Josephson junction
inserted in the central conductor~\cite{Niemczyk10a,Mallet09a}
indicate a feasibility of our scheme.

\begin{figure}
\includegraphics*[width=8cm]{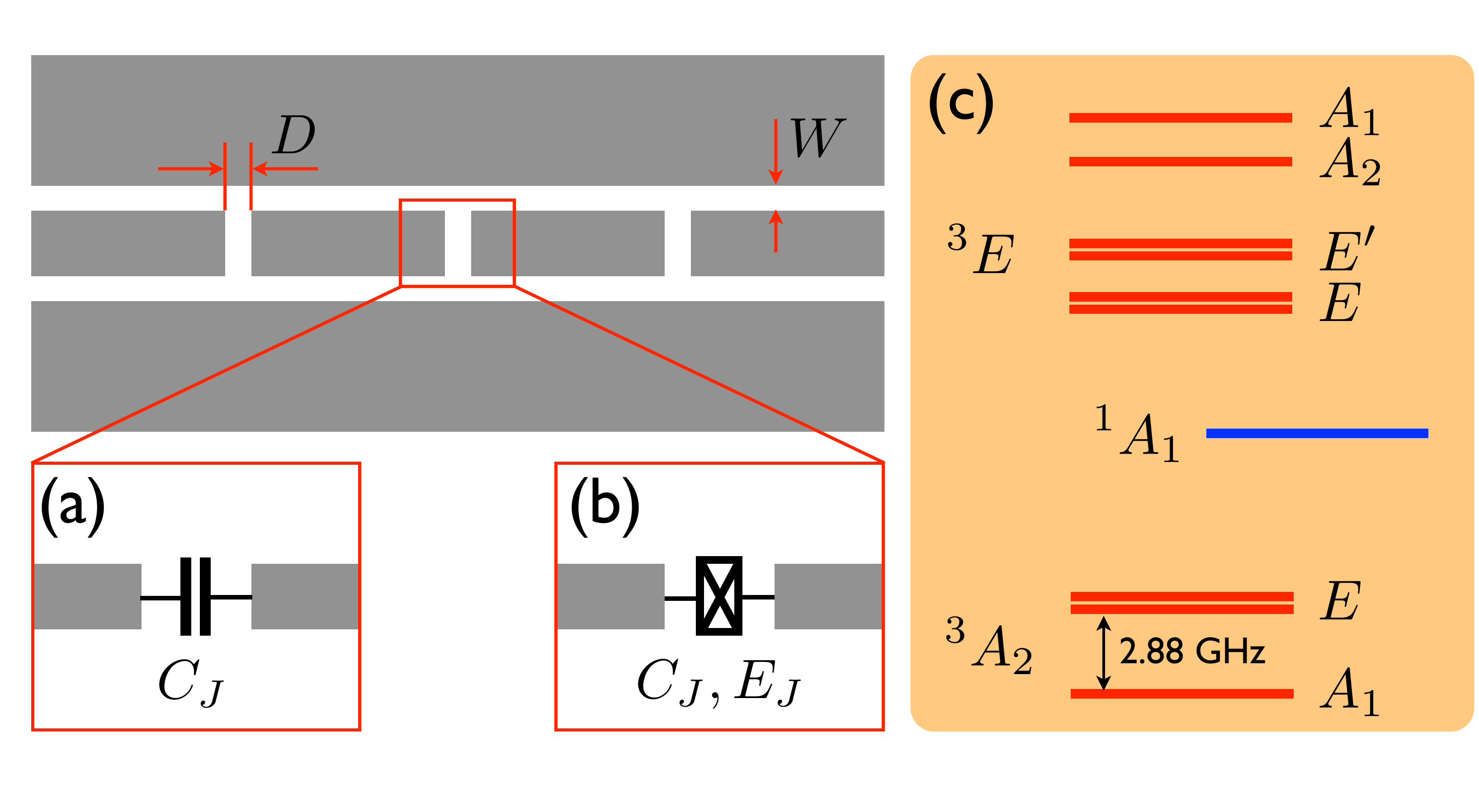}
\caption{(color online) A schematic diagram of the coplanar waveguide
  resonator with (a) capactive or (b) Josephson junctions. (c) The
  energy-level structure of the NV center in diamon.}
\label{paper::fig:1}
\end{figure}

\paragraph{Resonator.---}
We start with a general description of the resonator.  We insert $N$ junctions
into the central conductor of the superconducting coplanar waveguide,
essentially breaking the resonator into $N+1$ subresonators.  The adjacent
subresonators are coupled via the junctions.
The central conductor of the $r$th subresonator occupies the space $x_{2r}\leq
x\leq x_{2r+1}$ with length $\ell_r\equiv x_{2r+1}-x_{2r}$
($r=0,1,2,\cdots,N$) and has capacitance $C_0$ per unit length.
The junctions separating the adjacent subresonators have gap
$D_r=x_{2r}-x_{2r-1}$ ($r=1,\dots,N$) and junction capacitance
$C_J$.  We will assume that $\ell_r\sim 10\unit{mm}$ and $D_r\approx
1$--$10\unit{nm}$.

 The Lagrangian $\calL_R$ governing the dynamics
of the resonator has two parts: $\calL_R=\calL_0+\calL_J$.  The
first part $\calL_0$ describes the decoupled subresonators and is
given by~\cite{Blais04a}
\begin{equation}
\label{paper::eq:1}
\calL_0
= \half C_0\sum_{r=0}^{N}\int_{x_{2r}}^{x_{2r+1}}{dx}\,
\left[(\partial_t\phi)^2
  - v^2(\partial_x\phi)^2\right]
\end{equation}
where $v\approx 10^8\unit{m/s}$ is the propagating velocity of the
electromagnetic wave in the coplanar waveguide.  Physically, the field
$\phi(x,t)$ is proportional to the magnetic flux and related to the local
electric potential $V(x,t)$ by
\begin{math}
\partial_t\phi(x,t)=V(x,t)
\end{math}
The second part $\calL_J$ comes from the coupling between adjacent
subresonators:
\begin{equation}
\label{paper::eq:2}
\calL_J = \half C_J\sum_{r=1}^N
\left[(\partial_t\varphi_r)^2- \omega_p^2(\varphi_r)^2\right]
\end{equation}
Here the variable $\varphi_r(t)$ is the magnetic flux within the junction and
related to the electric potential difference $V_r(t)$ across the junction by
\begin{math}
\partial_t\varphi_r(t) = V_r(t) \,.
\end{math}
The first term in Eq.~(\ref{paper::eq:2}) is thus responsible for the
electric energy stored in the junction and the second, the magnetic
energy. $\omega_p=\sqrt{2E_CE_J}/\hbar\approx 2\pi\times
10\unit{GHz}$ is the Josephson plasma frequency, where
$E_C\equiv{}(2e)^2/2C_J$ and $E_J$ are the charging and Josephson coupling
energy of the junction, respectively.
The model~(\ref{paper::eq:2}) for Josephson junctions is valid only in the
range $k_BT\ll\hbar\omega_p\ll{}E_J$ whereas one can put $\omega_p=0$ in
(\ref{paper::eq:2}) for capacitive junctions.
Note that for large $D_r$ the
junction capacity $C_J$ (as well as the Josephson coupling $E_J$) becomes very
small compared with $\ell_rC_0\sim 1\unit{pF}$, and the coupling between
adjacent subresonators are negligible except for a small red-shift of order of
$C_J/\ell_rC_0$~\cite{Blais04a}. For $D_r\approx 1$--$10\unit{nm}$,
$C_J/\ell_rC_0\sim 1$ and the coupling is strong.

The field $\phi(x,t)$ defined inside subresonators and $\varphi_r(t)$ defined
across the junctions are not independent.  They are related to each other by
the current conservation
\begin{equation}
\label{paper::eq:10}
\frac{C_J}{v^2C_0}(\partial_t^2+\omega_p^2)\varphi_r(t)
= \partial_x\phi(x_{2r},t)
= \partial_x\phi(x_{2r-1},t)
\end{equation}
Following Ref.~\cite{Bourassa09a}, we expand the field $\phi$ and
$\varphi_r$ in the normal modes as
\begin{equation}
\label{paper::eq:11}
\phi(x,t) = \sum_{m=0}^\infty\phi_m(t)\psi_m(x) \,,\;
\varphi_r(t) = \sum_{m=0}^\infty\phi_m(t)\Delta_{r,m}.
\end{equation}
The eigenfunctions $\psi_m(x)$ should satisfy the time-independent
Schr\"odinger equation
\begin{equation}
\label{paper::eq:6}
(\partial_x^2+k_m^2)\psi_m(x)=0 \quad (x_{2r}\leq x\leq x_{2r+1})
\end{equation}
for some wave numbers $k_m$.  Let $\omega_m\equiv vk_m$. The current
conservation relation~(\ref{paper::eq:10}) now reads
\begin{equation}
\label{paper::eq:7}
\frac{C_J}{v^2C_0}(\omega_p^2-\omega_m^2)\Delta_{r,m}
= \partial_x\psi_m(x_{2r})
= \partial_x\psi_m(x_{2r-1})
\end{equation}
Without loss of generality, we choose the
normalization
\begin{equation}
\label{paper::eq:8}
\sum_{r=0}^N\int_{x_{2r}}^{x_{2r+1}}{dx}\,\psi_m\psi_n
+ C_J\sum_{r=1}^N\Delta_{r,m}\Delta_r\psi_n
= C_\Sigma\delta_{mn} \,,
\end{equation}
where
\begin{math}
C_\Sigma=\sum_{r=0}^N\ell_rC_0+NC_J
\end{math}
is the total capacitance of the resonator.
Putting the normal mode expansion~(\ref{paper::eq:11}) into
Eqs.~(\ref{paper::eq:1}) and (\ref{paper::eq:2}) and imposing the
conditions~(\ref{paper::eq:6}), (\ref{paper::eq:7}), and (\ref{paper::eq:8}),
one can rewrite $\calL_R$ into the simple form
\begin{equation}
\calL_R = \half C_\Sigma\sum_m
\left[(\partial_t\phi_m)^2 - \omega_m^2\phi_m^2\right]
\end{equation}
By introducing a momentum $\theta_m=C_\Sigma\partial_t\phi_m$ conjugate to
$\phi_m$, we write the Hamiltonian of the resonator as
\begin{equation}
H_R = \half\sum_m\left(\frac{\theta_m^2}{C_\Sigma}
  + C_\Sigma\omega_m^2\phi_m^2\right)
\end{equation}
We then quantize it by the canonical commutation relation
\begin{math}
[\phi_m,\theta_m] = i\hbar.
\end{math}
It is customary to introduce the annihilation and creation operators
of the normal modes by the relations
\begin{subequations}
\label{paper::eq:12}
\begin{align}
\phi_m &= \sqrt{\frac{\hbar}{2 \omega_m C_\Sigma}} (a_m^\dagger+a_m),\\
\theta_m &= i \sqrt{\frac{\hbar\omega_m C_\Sigma}{2}} (a_m^\dagger-a_m),
\end{align}
\end{subequations}
in terms of which the resonator Hamiltonian
reads
\begin{equation}
H_R = \sum_{m=0}^\infty\hbar\omega_ma_m^\dag a_m
\end{equation}
Due to the nonlinearity induced by the junctions,
$\omega_m$ determined by (\ref{paper::eq:7}) is not an integer
multiple of the $\omega_0$.

\paragraph{Coupling to NV centers.---}
We now describe the diamond NV centers and their coupling to the
resonator. The ground-state triplet (spin 1) of the NV center has a
level splitting $\epsilon/2\pi=2.88\unit{GHz}$ due to the spin-spin
interaction~\cite{Manson06a,Tamarat08a}, and thus is described by
the Hamiltonian
\begin{equation}
\label{paper::eq:3}
H_\mathrm{NV} = \hbar\epsilon\sum_{r=1}^NS_{r,z}^2
\end{equation}
where $S_{r,z}$ is the spin $z$ component of the NV center at the $r$th
junction.

The ground-state spin triplet is coupled magnetically to the resonator,
which is governed by the coupling Hamiltonian
\begin{equation}
\label{paper::eq:4}
H_g = \sum_{r=1}^Ng_e\mu_BS_{r,z}B_r
\end{equation}
where $B_r\equiv\varphi_r/D_rW$ ($W$ is the distance between the
central conductor and the ground plates) is the local magnetic field
at the junction $r$, $g_e\approx-2$ is the electron $g$-factor, and $\mu_B$
is the Bohr magneton.

Using the annihilation and creation operator defined in
Eq.~(\ref{paper::eq:12}), the coupling Hamiltonian is written as
\begin{equation}
H_g = \sum_{m,r}g_{m,r}(a_m^\dag+a_m)S_{r,z}
\end{equation}
Here we have defined the coupling constant
\begin{equation}
\label{paper::eq:13}
g_{m,r} = g_e\mu_B B_{m,r}
\end{equation}
where $B_{m,r}$ is the root-mean square value of the local magnetic
field in the mode $m$ at the junction $r$.  By
adjusting one of the mode frequencies at resonance with the level
splitting of the NV centers ($\omega_m\approx\epsilon$), one can
selectively couple the mode $m$ to the NV centers.

Putting all together, the total Hamiltonian is given by the sum
$H=H_R+H_{NV}+H_g$.
The description above is completely general (in principle) for any
number of junctions and NV centers.
Equations~(\ref{paper::eq:6}) and (\ref{paper::eq:7}) determines the possible
spectrum $\omega_m$ ($m=0,1,2,\cdots$) of the coupled subresonators whereas
the normalization~(\ref{paper::eq:8}) determines the magnitude of the magnetic
field at the junctions.
Below we demonstrate the cases with single NV center and two NV centers.

\begin{figure}
\centering
\includegraphics[width=4cm]{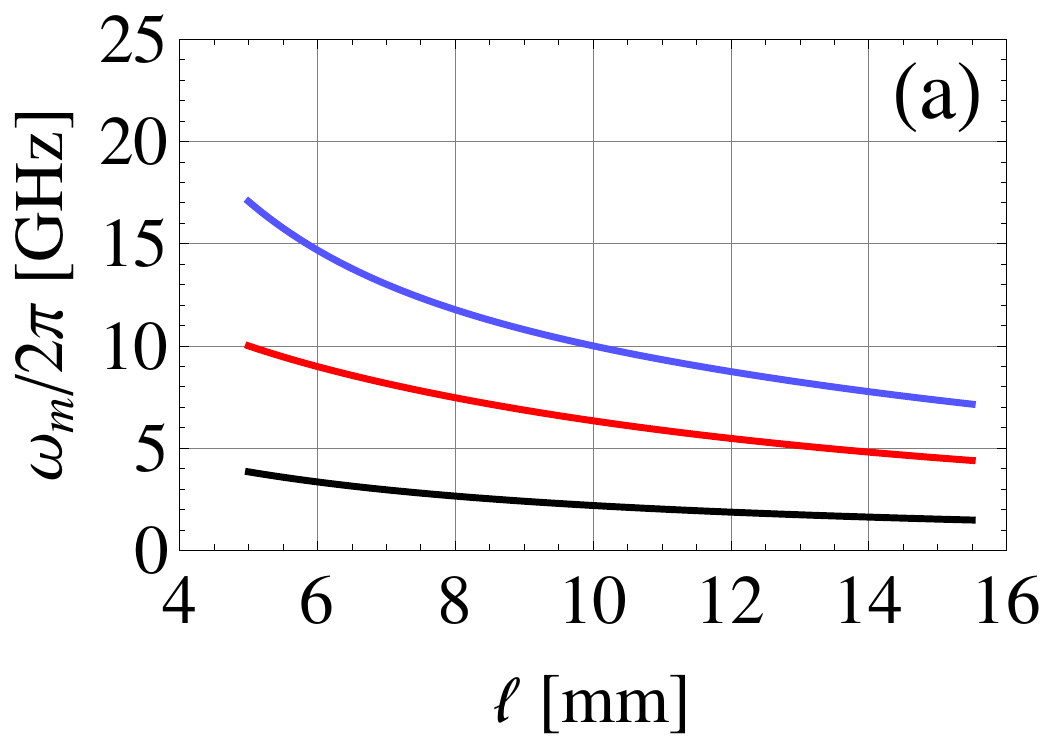}
\includegraphics[width=4cm]{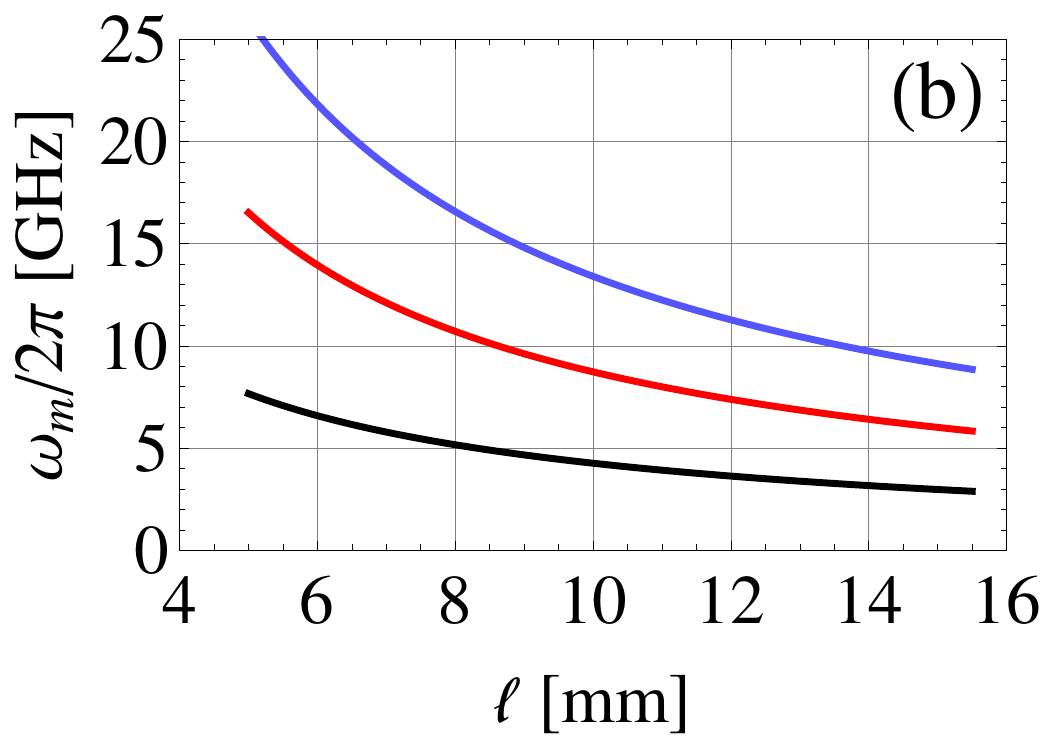}
\includegraphics[width=4cm]{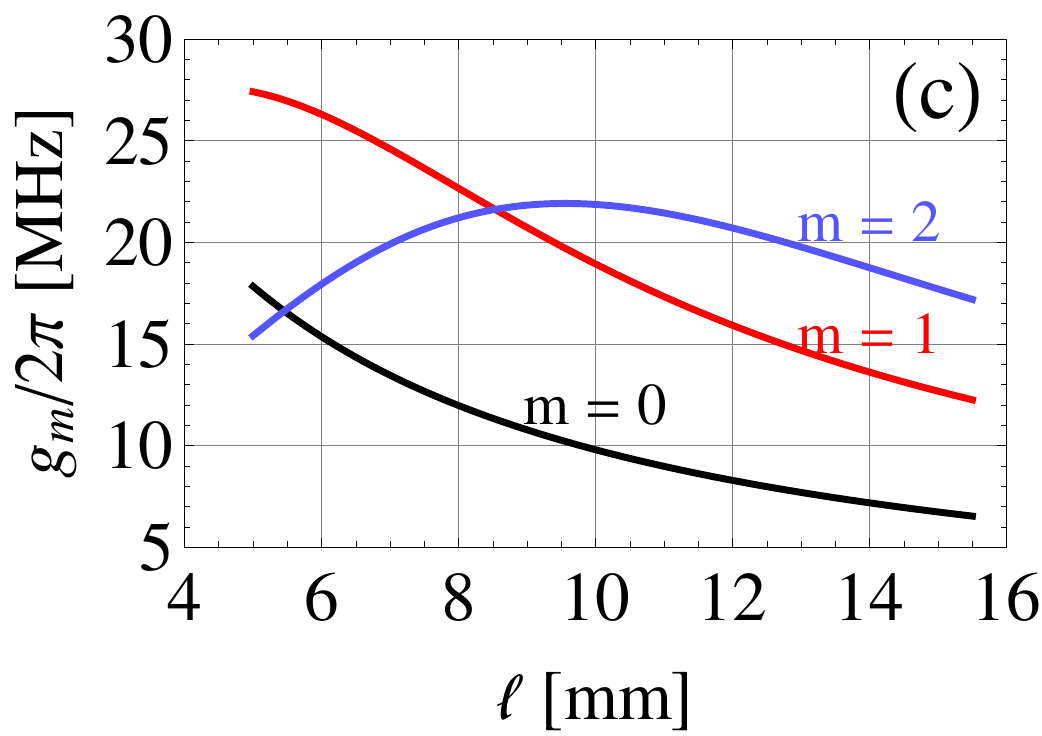}
\includegraphics[width=4cm]{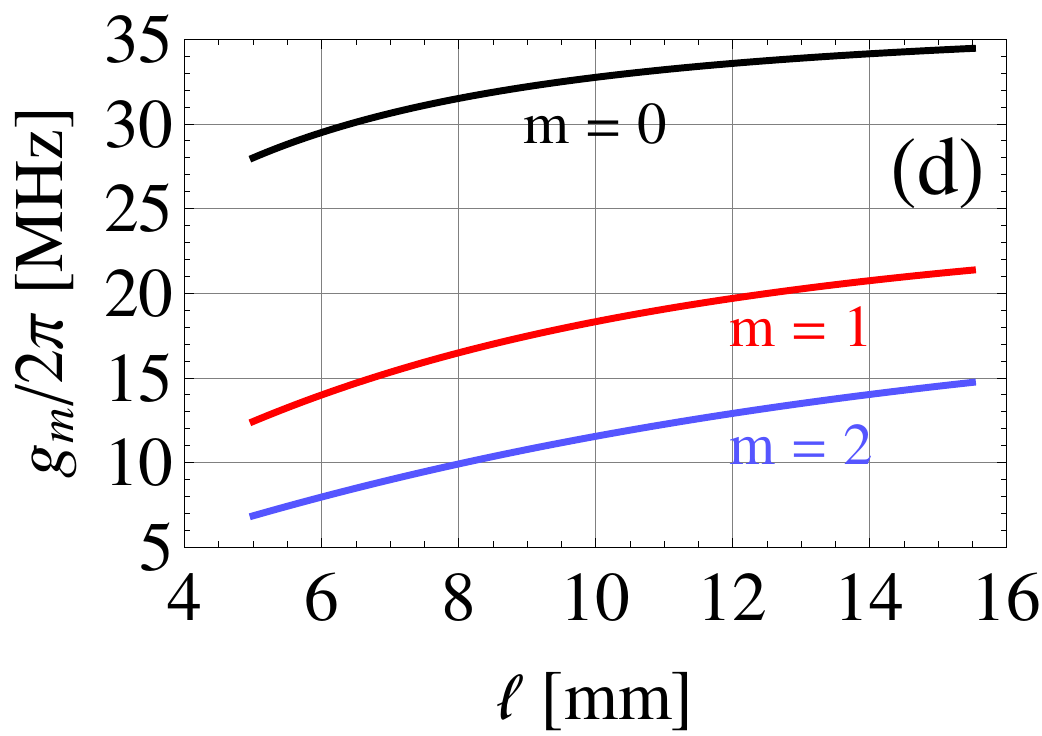}
\caption{(color online) (a,b) Resonance frequencies $\omega_m$ of the modes
  $m=0,1,2$ (from bottom to top) as a function of the overall length $\ell$ of
  the resonator for $\omega_p/2\pi=10\unit{GHz}$ (a) and $\omega_p=0$ (b).
  (c,d) The coupling strength $g_m$ to the modes $m=0,1,2$ for
  $\omega_p/2\pi=10\unit{GHz}$ (c) and $\omega_p=0$ (d). $C_J=0.15\unit{pF}$,
  $C_0=0.16\unit{pF/mm}$, $D=5\unit{nm}$, and $W=10\unit{\mu m}$.}
\label{paper:fig:2}
\end{figure}

\paragraph{Single NV center.---}
We assume that the junction is located at $x=0$, and the two subresonators are
identical with length $\ell$.  The wave function $\psi_m(x)$ of the normal
model $m$ has the form\cite{Bourassa09a}
\begin{equation}
\psi_m(x) = A_m
\begin{cases}
+\cos[k_m(x+\ell)] & (x\leq 0) \\
-\cos[k_m(x-\ell)] & (x>0) \,.
\end{cases}
\end{equation}
The wave number $k_m$ and the frequency $\omega_m=vk_m$ are
determined by the current conservation relation~(\ref{paper::eq:7}), which in
this case reduces to
\begin{equation}
\frac{2C_J}{v^2C_0}
\left(\omega_p^2 - \omega_m^2\right)
\cot[k_m \ell ] =  k_m,
\end{equation}
and the constant $A_m$ (whose explicit expression is not given) is
determined by the normalization condition~(\ref{paper::eq:8}).
The coupling strength $g_m$ in
Eq.~(\ref{paper::eq:13}) takes the form
\begin{equation}
g_m = \frac{4\mu_BA_m\cos(k_m\ell)}{DW} \sqrt{\frac{\hbar}{2\omega_m
C_\Sigma}}
\end{equation}
Figure~\ref{paper:fig:2} shows $\omega_m$ and $g_m$ as a function of
the length $\ell$ for a few lowest modes. It demonstrates that the
coupling strength is in the strong coupling regime
($g_0/2\pi\sim10\unit{MHz}$) when the resonator is close to
resonance to the diamond NV center ($\omega_m\approx\epsilon$).
Notice that a capacitive junction gives coupling as strong
as a Josephson junction. A submicron size of the gap junction has
been made on coplanar waveguide~\cite{Ketterl05a}.

\begin{figure}
\centering
\includegraphics*[width=4cm]{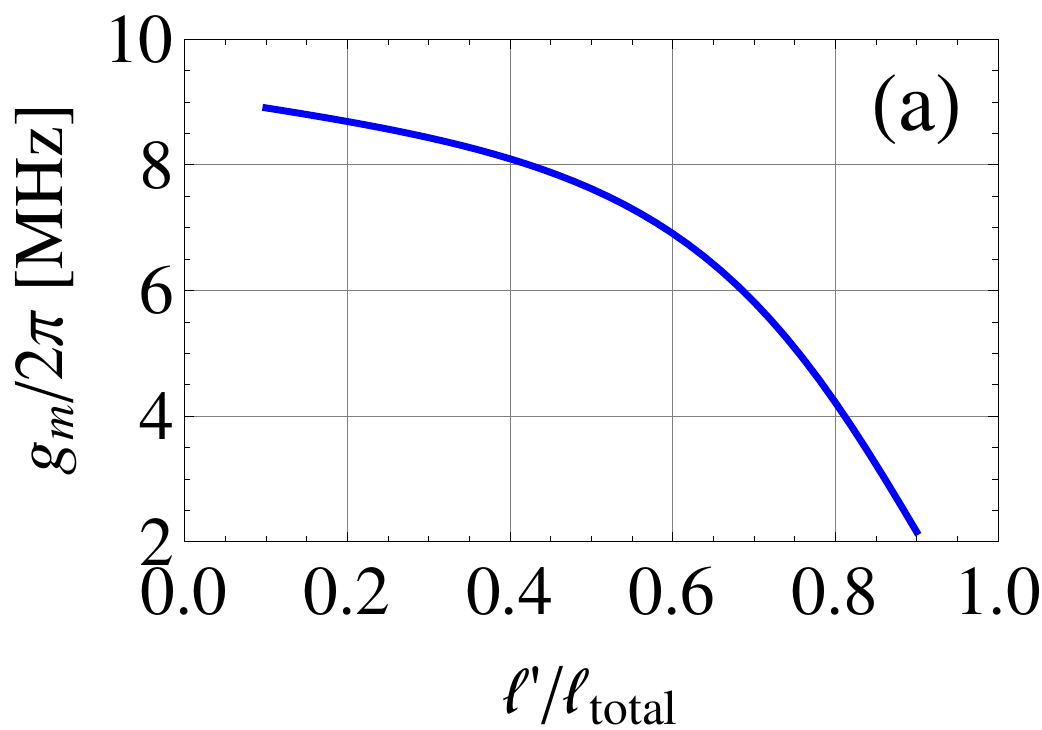}
\includegraphics*[width=4cm]{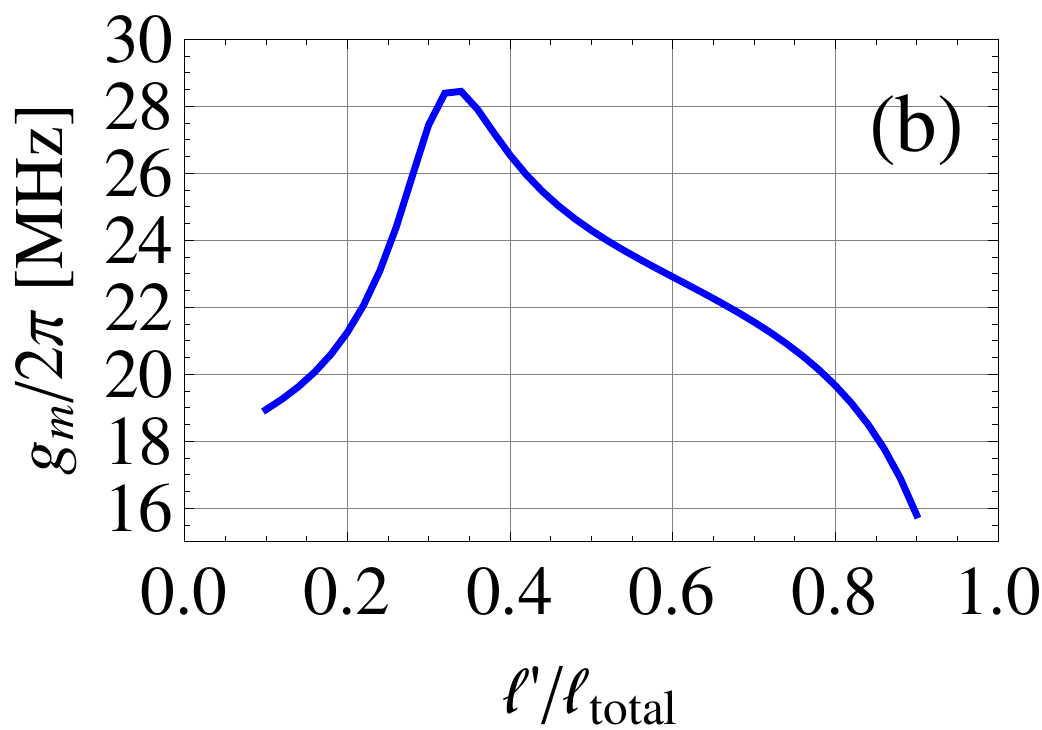}
\caption{The coupling strength (a) for $\omega_p/2\pi=10\unit{GHz}$ and
  $\ell_\mathrm{total}=18\unit{mm}$ and (b) for $\omega_p/2\pi=0$ and
  $\ell_\mathrm{total}=25\unit{mm}$. $C_J=0.15\unit{pF}$,
  $C_0=0.16\unit{pF/mm}$, $D=5\unit{nm}$, and $W=10\unit{\mu m}$.}
\label{paper::fig:3}
\end{figure}

\paragraph{Double NV center.---}
Let us now consider the case with two NV centers.  For simplicity,
we assume symmetrically located NV centers with subresonators of
lengths $\ell,\ell',\ell$ in this order ($x_5=-x_0=\ell+\ell'/2$,
$x_4=-x_2=\ell'/2$, and $D=x_2-x_1=x_4-x_3\ll\ell,\ell'$).  Because
of the symmetry, a normal-mode wavefunction $\psi_m$ has either even
or odd parity.
Here, we only consider the lowest mode, i.e., the $\lambda/2$ mode. The
eigenfunction takes the form of
\begin{equation}
\psi(x) =A
\begin{cases}
\frac{\sin(k\ell)}{\cos(k\ell'/2)}\sin(kx) &
(0\leq x\leq\ell'/2) \\
\cos k (x-\ell-\ell'/2) & (\ell'/2\leq x\leq\ell)
\end{cases}
\end{equation}
The mode frequency $\omega=vk$ is determined by the current
conservation~(\ref{paper::eq:7})
\begin{equation}
vC_0\omega=C_J\left(\omega_p^2-{\omega}^2\right)
[\cot(k\ell)-\tan(k\ell'/2)]
\end{equation}
while the constants $A$ (not given here) is determined from the
normalization condition~(\ref{paper::eq:8}). The coupling strength at the
junctions is found to be
\begin{align}
g=\frac{2\mu_BA}{WD}\sqrt{\frac{\hbar}{2C_\Sigma \omega}}
\left[\cos(k\ell) - \sin(k\ell)\tan(k\ell'/2)\right]
\end{align}
The coupling strength $g$ as a function of $\ell'$ is shown
in Fig.~(\ref{paper::fig:3}).

Let us now estimate the coupling strength $J$ between two NV
centers through a virtual excitation of the photon in a dispersive regime.
For the resonators coupled through Josephson junctions, we have the frequency,
$\omega/2\pi=2.4\unit{GHz}$, and the NV center-resonator coupling,
$g/2\pi\sim8\unit{MHz}$, for
$\ell_\mathrm{total}\equiv2\ell+\ell'=18\unit{mm}$ and $\ell'\sim0.4\unit{mm}$.
Assuming a detuning of $\Delta/2\pi=200\unit{MHz}$ between $\epsilon$ and
$\omega$, we obtain the coupling strength between the NV centers
$J=g^2/\Delta\sim2\pi\times320\unit{kHz}$. For $Q=10^5$, the cavity decay rate
is $\kappa=\omega/Q\sim2\pi\times20\unit{kHz}$ which is lower than the
coupling strength $J$. Even for $J\sim\kappa$, two-qubit quantum gates can be
performed with a high-fidelity as demonstrated in Ref.~\cite{DiCarlo09a}.
Thus, given the spin coherence time of
$1-10\unit{ms}$~\cite{Balasubramanian09a},
a high fidelity quantum gate is realizable using our scheme.

\textit{Discussion} -- Above we have focused on how to strongly couple diamond
NV centers to the resonator.  The strong coupling opens another interesting
possibility to couple NV centers to other types of superconducting qubits
\cite{Wallraff04a,Blais04a,Majer07a,Deppe08a,Imamoglu09a,Bourassa09a,Niemczyk10a}
through the resonator. This way one can take the best of features that each
qubit provides. It will ultimately provide a novel architecture for quantum
information processors integrating diamond NV centers and superconducting
qubits into the circuit QED system.


\begin{acknowledgments}
This work was supported by the NRF Grant 2009-0080453 (MEST Korea), the BK21,
and the APCTP. M.-S.C thanks R. Aguado and D. Marcos for their useful
discussions and comments.
\end{acknowledgments}

\bibliographystyle{apsrev}
\bibliography{paper}

\end{document}